\documentclass[12pt]{article}
\usepackage{amsmath,amsthm,amsfonts}
\usepackage{url}
\newcommand{\pf}{\noindent {\bf Proof:} }
\DeclareMathAlphabet{\mathpzc}{OT1}{pzc}{m}{it}
\newtheorem{theorem}{Theorem}

\newtheorem{lemma}[theorem]{Lemma}

\title{Patterns with involutions}
\author{James Currie\thanks{The author is
supported by an NSERC Discovery Grant.}\\
Department of Mathematics and Statistics \\
University of Winnipeg \\
515 Portage Avenue \\
Winnipeg, Manitoba R3B 2E9 (Canada) \\
\url{j.currie@uwinnipeg.ca} }
\begin{document}
\date{\today}
\maketitle
\begin{abstract}
\noindent \title{We give the avoidance indices for all unary patterns with involution.}
\vspace{.1in}\\
\noindent Keywords: words avoiding patterns, combinatorics on words, repetitions, Thue-Morse word
\end{abstract}
Consider a non-empty word $p$ over $\Sigma=\{x, g(x)\}.$ Here $g(x)$ is literally the string $`g(x)$', so that if $p=xg(x)x$, we say that $|p|=3$. Let $T$ be a finite alphabet. We call $w\in T^*$ a {\bf morphic instance} (resp., {\bf antimorphic instance}) of $p$ if there is a morphic (resp,, antimorphic) involution $g_T$ of $T^*$ and a non-erasing morphism $\phi:\Sigma^*\rightarrow T^*$ such that $\phi( g(x))=g_T(\phi(x)).$ In the case that $p=xg(x)x$, a morphic (resp., antimorphic) instance of $w$ would be a word $yg_T(y)y$ where $y\in T^+$ and $g_T$ is a morphic (resp., antimorphic)  involution of $T^*$. The {\bf morphic (resp., antimorphic) avoidance index of $p$} is the size of the smallest alphabet $T$ such that there exists an infinite word over $T$, no factor of which is a morphic (resp., antimorphic) instance of $p$. Denote the morphic (resp., antimorphic) avoidance index of $p$ by $A_m(p)$ (resp., $A_a(p)$). If the symbol $g(x)$ doesn't appear in $p$, then $A_m(p)=A_a(p)$ is just the usual avoidance index of $p$. As is pointed out in \cite{B&M}, interchanging $x$'s and $g(x)$'s in a pattern $p$ does not change the morphic or antimorphic avoidance index. If $p\in \Sigma^3$, the avoidance index of $p$ is given in \cite{B&M}:
$$A_m(p)=A_a(p) =\left\{\begin{array}{ll}
2,&p\in\{xxx,g(x)g(x)g(x)\}\\
3&\mbox{ otherwise}
\end{array}\right.$$

Although $\Sigma$ has two elements, it is natural to call words over $\Sigma$ {\bf unary patterns with involution}. The next most complex patterns to consider would be over $\{x,y,g(x),g(y)\}$, and we would consider them {\bf binary patterns with involution.} We will give the avoidance indices for all unary patterns with involution.

The avoidance indices of words $x^n$ are known, so we need only consider words $p$ for which $|p|_x,|p|_{g(x)}\ge 1.$ Clearly, $A_m(xg(x))=A_a(xg(x))=\infty.$ The avoidance indices for patterns of length 3 are known.  We will show that whenever $p\in\Sigma^4$, $A_m(p)=A_a(p)=2$. Since no word can have avoidance index 1, we see that

$$A_m(p)=A_a(p) =\left\{\begin{array}{ll}
3,&p\in\Sigma^3-\{xxx,g(x)g(x)g(x)\}\\
\infty,&p\in\{x,g(x),xg(x),g(x)x\}\\
2&\mbox{ otherwise}
\end{array}\right.$$

The avoidance indices are clearly 2 when $xxx$ is a factor of $p$. We consider words $p\in\Sigma^4$ where $xxx$ is not a factor. Interchanging $x$'s and $g(x)$'s if necessary, assume that $|p|_x\ge |p|_{g(x)}$. Since $xxx$ is not to be a factor of $p$, either $|p|_{g(x)}=1$ or $|p|_{g(x)}=2$. In the first case, our word $p$ is $xxg(x)x$ or $xg(x)xx$. Since avoidance indices are preserved under reversal, we need only consider the case $p=xxg(x)x$ here. If $|p|_{g(x)}=2$, ignoring reversals, we consider $xg(x)xg(x),g(x)xxg(x),xxg(x)g(x)$. For each of these $p\in\Sigma^4$ we will show that both avoidance indices are 2. Simplifying (or abusing, if you prefer) our notation, this amounts to constructing an infinite binary word with no factor $xxg(x)x$ ($xg(x)xg(x),g(x)xxg(x),xxg(x)g(x)$) where $x$ is non-empty and $g$ is a morphic ($g$ is an antimorphic) involution.

\section{Morphic involutions}
Let {\bf t} be the Thue-Morse sequence $h^\omega(0)$, where $h(0)=01$, $h(1)=10$.
Write ${\bf t}=\Pi_{i=0}^\infty t_i$, $t_i\in\{0,1\}$.

Let ${\bf w}$ be the infinite word $${\bf w}=\Pi_{j=0}^\infty 0^21^{t_i+2}.$$
We see that {\bf w} is concatenated from blocks of two 0's alternated with blocks of either two or three 1's.
\begin{lemma}\label{xxXx}Word {\bf w} has no factor of the form $xxg(x)x$ where $x$ is a non-empty word and $g(x)$ is the image of $x$ under a morphic involution of $\{0,1\}^*$.
\end{lemma}
\pf Suppose for the sake of getting a contradiction that $xxg(x)x$ is a factor of {\bf w}
where $x$ is a non-empty word and $g(x)$ is a morphic involution of $\{0,1\}^*$.

If $|x|_0=0$, then $x=1^m$ for some $m$. If $g$ is the identity, this makes 1111 a factor of {\bf w}, which is impossible. If $g$ is the complement morphism, then $m\le 2$, since $g(x)=0^m$ is a factor of {\bf w}. Then, however, $xxg(x)x=1101$ or 11110011, neither of which is a factor of {\bf w}. If $|x|_1=0$, then $x=0$ or $x=00$. If $g$ is the identity, this makes 0000 a factor of {\bf w}, which is impossible. If $g$ is the complement morphism, then 0010 or 00001100 is a factor of {\bf w} neither of which is possible.
We conclude that $|x|_0,|x|_1\ge 1$.

Suppose that $g$ is the complement morphism. Word {\bf w} has factors $g(x)x$ and $xx$, hence factors $0x$, $1x$. This means that $x$ cannot start $01$, 10 or 00, since none of 101, 010 or 000 are factors of {\bf w}. We deduce that $x$ commences 11. Similarly, $x$ ends 11. Now, however, $xx$ has 1111 as a factor, which is impossible.

Suppose then that $g$ is the identity morphism, so that $xxxx$ is a factor of {\bf w}. Let $s\ge 0$ be maximal so that $0^s$ is a prefix of $x$. Let $t\ge 0$ be maximal such that $0^t$ is a suffix of $x$. Since $|x|_1\ge 1$, $x$ has prefix $0^s1$ and suffix $10^t$, and $10^{t+s}1$ is a factor of $xx$, implying $t+s=0$ or $t+s=2$.

\noindent{\bf Case 1: } Suppose $t+s=0$. If $|x|_0=2$, write $x=1^r0^21^q$, $r,q\ge 1$. Then $xxxx=1^r0^21^{q+r}0^21^{q+r}0^21^{q+r}0^21^q$, and {\bf t} contains the overlap $(q+r-2)(q+r-2)(q+r-2)$, which is impossible. Thus assume $|x|_0>2$, and write $x = 1^r0^21^{t_i+2}0^2\cdots1^{t_j+2}0^21^q$, $r,q\ge 1$, $i\le j$. Then $xxxx$ is $$1^r0^21^{t_i+2}\cdots1^{t_j+2}0^21^{q+r}0^21^{t_i+2}\cdots1^{t_j+2}0^21^{q+r}0^21^{t_i+2}\cdots1^{t_j+2}0^21^{q+r}0^21^{t_i+2}0^2\cdots1^{t_j+2}0^21^q,$$ and {\bf t} contains the overlap $(q+r-2)t_i\cdots t_j(q+r-2)t_i\cdots t_j(q+r-2)$, which is again impossible.

\noindent{\bf Case 2: }Suppose $t+s=2$. If $|x|_0=2$, write $x=0^s1^{t_i+2}0^t$, some $i$. Then $xxxx=0^s1^{t_i+2}0^21^{t_i+2}0^21^{t_i+2}0^21^{t_i+2}0^t$, and {\bf t} contains the overlap $t_it_it_i$, which is impossible. Thus assume $|x|_0>2$, and write $x = 0^s1^{t_i+2}0^2\cdots1^{t_j+2}0^t$, $i\le j$. Then $xxxx$ is $$0^s1^{t_i+2}\cdots1^{t_j+2}0^21^{t_i+2}\cdots1^{t_j+2}0^21^{t_i+2}\cdots1^{t_j+2}0^21^{t_i+2}\cdots1^{t_j+2}0^t,$$ and {\bf t} contains the overlap $t_i\cdots t_jt_i\cdots t_jt_i$, which is again impossible.$\Box$

Let ${\bf v}$ be the infinite word $${\bf v}=\Pi_{j=0}^\infty 01^{2t_i+1}.$$
We see that {\bf v} is concatenated from 0's alternated with blocks of either one or three 1's.
\begin{lemma}\label{XxxX}Word {\bf v} has no factor of the form $g(x)xxg(x)$ where $x$ is a non-empty word and $g(x)$ is the image of $x$ under a morphic involution of $\{0,1\}^*$.
\end{lemma}
\pf Suppose for the sake of getting a contradiction that $g(x)xxg(x)$ is a factor of {\bf v}
where $x$ is a non-empty word and $g(x)$ is a morphic involution of $\{0,1\}^*$.

Since 00 is not a factor of {\bf v} but $xx$ is a factor, $|x|_1\ge 1$. If $|x|_0=0$, then $x=1^m$ for some $m$. If $g$ is the identity, this makes 1111 a factor of {\bf v}, which is impossible. If $g$ is the complement morphism, then $m=1$, since $g(x)=0^m$ is a factor of {\bf v}. Then, however, $g(x)xxg(x)=0110$, which is not a factor of {\bf v}. We conclude that $|x|_0,|x|_1\ge 1$.

Suppose that $g$ is the complement morphism. If $x$ begins and ends with different letters, then one of $g(x)x$ and $xg(x)$ has 00 as a factor, which is impossible. Therefore the first and last letters of $x$ are the same. They must both be 1; otherwise $xx$ would contain 00. Again 11 cannot be a factor of $x$; otherwise 00 would be a factor of $g(x)$. It follows that $x$ begins with 10 and ends with 01. Now, however, $xx$ has the factor 0110, which is impossible.

Suppose then that $g$ is the identity, so that $xxxx$ is a factor of {\bf v}.
If $|x|_0=1$, write $x=1^q01^r$, some $q,r\ge 0$. We must have $q+r\ge 1$, since $|x|_1\ge 1$.
Now $xxxx=1^q01^{r+q}01^{r+q}01^{r+q}01^t$.  This implies the existence of an overlap $\frac{r+q-1}{2}\frac{r+q-1}{2}\frac{r+q-1}{2}$ in {\bf t}, which is impossible. 

Assume then that $|x|_0\ge 2$. Write $x=1^q01^{2t_i+1}\cdots 1^{2t_j+1}01^r$ for some $i\le j$, some $q,r\ge 0$. Then $xxxx$ has the factor
$$1^{r+q}01^{2t_i+1}\cdots 1^{2t_j+1}01^{r+q}01^{2t_i+1}\cdots 1^{2t_j+1}01^{r+q}0$$
and {\bf t} contains the overlap $$\frac{r+q-1}{2}t_i \cdots t_j\frac{r+q-1}{2}t_i\cdots t_j\frac{r+q-1}{2}.$$
This is impossible. $\Box$


 Let ${\bf u}$ be the infinite word $${\bf u}=\Pi_{j=0}^\infty 01^{t_i+2}.$$
We see that {\bf u} is concatenated from 0's alternated with blocks of either 3 or 2 1's.
\begin{lemma}\label{xxXX}Word {\bf u} has no factor of the form $xxg(x)g(x)$ or $xg(x)xg(x)$ where $x$ is a non-empty word and $g(x)$ is the image of $x$ under a morphic involution of $\{0,1\}^*$.
\end{lemma}
\pf Suppose for the sake of getting a contradiction that $xxg(x)g(x)$ or $xg(x)xg(x)$ is a factor of {\bf u}
where $x$ is a non-empty word and $g(x)$ is a morphic involution of $\{0,1\}^*$.

First suppose that $g$ is the complement morphism. Since {\bf u} contains a factor $g(x)$, but no factor 00, word $x$ cannot contain 11 as a factor. Similarly, {\bf u} doesn't contain a factor 010, so that $x$ cannot contain a factor 101. The only possibilities for $x$ are then 0, 1, 01 and 10. The resulting values for $xxg(x)g(x)$ (resp. $xg(x)xg(x)$) would be 0011, 1100, 01011010, 10100101 (resp. 0101, 1010, 01100110, 10011001) which all contain either 00 or 010 and are thus impossible. 

Suppose then that $g$ is the identity morphism. Thus $xxg(x)g(x)=xg(x)xg(x)=xxxx.$
Since 00 is not a factor of {\bf u} but $xx$ is a factor, $|x|_1\ge 1$. If $|x|_0=0$, then $x=1^m$ for some $m$, and 1111 is a factor of {\bf u}. This is impossible. It follows that $|x|_0,|x|_1\ge 1$. If $|x|_0=1$, write $x=1^q01^r$, some $q,r\ge 0$.
Then $xxxx=1^q01^{r+q}01^{r+q}01^{r+q}01^t$.  This implies the existence of an overlap $(r+q-2)(r+q-2)(r+q-2)$ in {\bf t}, which is impossible. 

Assume then that $|x|_0\ge 2$. Write $x=1^q01^{t_i+2}\cdots 1^{t_j+2}01^r$ for some $i\le j$, some $q,r\ge 0$. Then $xxxx$ has the factor
$$1^{r+q}01^{t_i+2}\cdots 1^{t_j+2}01^{r+q}01^{t_i+2}\cdots 1^{t_j+2}01^{r+q}$$
and {\bf t} contains the overlap $$(r+q-2)t_i\cdots t_j(r+q-2)t_i\cdots t_j(r+q-2).$$
This is impossible. $\Box$

\section{Antimorphic involutions}

Over $\{0,1\}$, there are only two antimorphisms: the reversal $x\rightarrow x^R$ generated by $0^R=0$ and $1^R=1$, and the reverse complement $x\rightarrow \bar{x}^R$.

\begin{lemma}Word {\bf w} has no factor of the form $xxg(x)x$ where $x$ is a non-empty word and $g(x)$ is the image of $x$ under an antimorphic involution of $\{0,1\}^*$.
\end{lemma}
\pf Suppose for the sake of getting a contradiction that $xxg(x)x$ is a factor of {\bf w}
where $x$ is a non-empty word and $g(x)$ is an antimorphic involution of $\{0,1\}^*$. 

By Lemma~\ref{xxXx} we may assume that $g(x)\ne x$, since we have shown that {\bf w} has no factor $xxxx$ with $x$ non-empty. Similarly, we may assume that $g(x)\ne \bar{x}.$ These conditions together imply that $x$ is not a palindrome, and that $x^R\ne \bar{x}$. Suppose, for example, that $x$ is a palindrome. If $g$ is reversal, then $g(x)=x$, which we have forbidden. If $g$ is reverse complement, then $g(x)=\overline{(x^R)}=\bar{x}$, again forbidden. Similarly one checks that $x^R\ne \bar{x}$.
To continue with our proof, suppose that $g$ is the reverse complement. Since {\bf w} contains a factor $g(x)$, but no factor 000, word $x$ cannot contain 111 as a factor. Also, {\bf w} does not contain 010 or 101 as a factor. It follows that $x$ is a factor of $(0011)^\omega$. Since $xg(x)$ and $g(x)x$ are factors of {\bf w}, $x$ cannot begin or end with $01$ or $10$. It therefore begins and ends with 00 or 11. The length 2 prefix and length 2 suffix of $x$ must differ, since otherwise $xx$ would have 0000 or 1111 as a factor. We conclude that $x=(0011)^n$ or $x=(1100)^n$ for some $n$. But then $x$ is the complement of its reverse, contradicting our previous assumption.

Suppose then that $g$ is the reversal. Since $xg(x)$ and $xx$ are both factors of {\bf w} but 010, 101 are not, $x$ cannot end in 01 or 10. Then $x$ ends in 00 or 11, and $xg(x)$ contains 0000 or 1111 as a factor. This is impossible. $\Box$

\begin{lemma}Word $(0001)^\omega$ has no factor of the form $xxg(x)g(x)$, $xg(x)xg(x)$ or $g(x)xxg(x)$ where $x$ is a non-empty word and $g(x)$ is the image of $x$ under an antimorphic involution of $\{0,1\}^*$.
\end{lemma}
\pf Suppose for the sake of getting a contradiction that $xxg(x)g(x)$, $xg(x)xg(x)$ or $g(x)xxg(x)$ is a factor of $(0001)^\omega$
where $x$ is a non-empty word and $g(x)$ is an antimorphic involution of $\{0,1\}^*$.  

If $g$ is reversal, then $x$ cannot end in 01 or 10;  this would imply 0110 or 1001 as a factor of $xg(x)$; however these are not factors of $(0001)^\omega.$ It follows that if $|x|>1$ then $x$ ends in $00$, since 11 is not a factor of $(0001)^\omega$. Then, however 0000 is a factor of $xg(x)$, which is impossible. We conclude that $|x|=1$, and $xxg(x)g(x),xg(x)xg(x),g(x)xxg(x)\in \{1111, 0000\}$. This is impossible.

If $g$ is reverse complement, 00 cannot be a factor of $x$; otherwise 11 is a factor of $g(x)$. However, $x$ cannot end in 01 or 10, or $xg(x)$ would have 0101 or 1010 as a factor. We conclude that $|x|=1$, and   $xxg(x)g(x)=xg(x)xg(x)=g(x)xxg(x)\in\{0011,0101,1001\}$, which are impossible.$\Box$

 \end{document}